\documentstyle[12pt]{article}
\def\appendix#1{
\addtocounter{section}{-2}
\setcounter{equation}{0}
\renewcommand{\thesection}{\Alph{section}}
\section*{Appendix \thesection\protect\indent #1}
\addcontentsline{toc}{section}{Appendix \thesection\ \ \ #1}
}
\newcommand{\tr}[1]{\,{\rm tr}\,#1\,}

\def\e{\varepsilon}

\def\be{\begin{equation}}
\def\la{\label}
\def\ee{\end{equation}}
\def\bea{\begin{eqnarray}}
\def\eea{\end{eqnarray}}
\def\eps{\varepsilon}

\def\b{\beta}
\def\s {\sigma}
\def\n{\nabla}

\def\D{\Delta}
\def\G{\Gamma}

\def\d{\delta}

\def\l{\left(}
\def\r{\right)}
\def\p{\partial}

\begin{document}
\begin{titlepage}
\thispagestyle{empty}
\title{
\large{
\begin{flushright}
LMU-TPW 00-9\\
UAHEP 00-4\\
hep-th/0003116
\end{flushright} }
\vskip 0.5cm
{\bf \huge{ 
A note on gravity-scalar fluctuations in holographic RG flow geometries   
}}}  
\author{G. Arutyunov,$^{a\, c}$
\thanks{arut@theorie.physik.uni-muenchen.de} \mbox{}
\mbox{} S. Frolov$^{b\,c}$
\thanks{frolov@bama.ua.edu }
 and \mbox{} S. Theisen$^{a}$
\thanks{theisen@theorie.physik.uni-muenchen.de
\newline
$~~~~~$$^c$On leave of absence from 
Steklov Mathematical Institute, Gubkin str.8, GSP-1, 
117966, Moscow, Russia}
\vspace{0.4cm} \mbox{} \\
\small {$^a$ Sektion Physik,}
\vspace{-0.1cm} \mbox{} \\
\small {Universit\"at M\"unchen,}
\small {Theresienstr. 37,}
\vspace{-0.1cm} \mbox{} \\
\small {D-80333 M\"unchen, Germany}
\vspace{0.4cm} \mbox{} \\
\small {$^b$ Department of Physics and Astronomy,}
\vspace{-0.1cm} \mbox{} \\
\small {University of Alabama, Box 870324,}
\vspace{-0.1cm} \mbox{} \\
\small {Tuscaloosa, Alabama 35487-0324, USA}
\mbox{}
}            
\date {}
\maketitle
\begin{abstract}

We study five-dimensional gravity models with non-vanishing 
background scalar fields which are dual to non-conformal 
boundary field theories. We develop a procedure to decouple 
the graviton fluctuations from the scalar ones and apply 
it to the simplest case of one scalar field. The quadratic action 
for the decoupled scalar fluctuations 
has a very simple form and can be 
used to compute two-point functions. 
We perform this computation for the two examples
of background RG flow 
recently considered by DeWolfe and Freedman 
and find physically reasonable results.
\end{abstract}
\end{titlepage}
\newpage
\section{Introduction}
The AdS/CFT correspondence \cite{M,GKP,W} provides a powerful method 
for computing correlation functions of gauge invariant operators
in ${\cal N}=4$ supersymmetric Yang-Mills theory (YM) in four dimensions
at large $N$ and at strong 't Hooft coupling. A question of obvious
interest is to extend the correspondence and in particular the 
prescription for computing Green functions to a general class 
of theories without (or with spontaneously broken) superconformal invariance. 
Such theories have a renormalization group flow and they generically arise 
either from deforming YM away from the conformal point
by adding to the Lagrangian IR relevant 
perturbations \cite{GPPZ0}-\cite{KT} or 
by giving vacuum expectation values to scalar fields 
\cite{KLT}-\cite{CGLP}. 
In the gauged supergravity description the corresponding bulk geometries 
are supported in both cases by 4-dimensional Poincar\'e invariant 
kink solutions approaching asymptotically AdS space 
(see \cite{G} for a recent extensive review of the subject).

Recently a gravity computation of some 
two-point correlation functions 
in field theories with renormalization group 
flow was presented in \cite{DF}.
This computation amounts to the 
study of the coupled gravity-scalar equations describing 
fluctuations of the scalars and the metric around a kink 
solution. The background scalars naturally fall into two classes: 
the scalars with a non-trivial dependence on the fifth dimension 
which, in the terminology of \cite{DF},
are called ``active'' and the constant (or vanishing) ``inert'' scalars. 
In \cite{DF} striking differences between the expected correlation 
functions of YM operators dual to inert and to active scalars were observed. 
The correlation functions of the operators dual to the inert scalars 
are straightforward to compute and they provide a consistent description 
of the spectrum in the boundary field theory. 
Quite opposite, the correlation functions of the operators dual to active 
scalars were then found to be physically unreasonable. 
The authors of \cite{DF} suggested that this might be due  
to the presence of the metric singularity in the 
interior, which in turn might invalidate
the standard AdS prescription for computing correlation functions. 
Adding the standard supplementary gravity boundary 
terms did not clarify the situation.

In this note we reconsider 
the two-point correlation functions for operators dual to active scalars.
We use the Hamiltonian formulation of the AdS/CFT 
correspondence \cite{AF1} to deal with boundary terms and  
the standard prescription for 
computing correlation functions in 
the gravity approximation \cite{GKP,W}.
Another point of deviation from the analysis in \cite{DF} is our 
choice of gauge.

We start by analyzing the quadratic action for 
fluctuations of the scalar fields and the metric near their 
background values. All the boundary terms one could
add to the gravity action are unambiguously fixed 
by the Hamiltonian version of the AdS/CFT correspondence \cite{AF1} 
that requires the gravity action to be schematically of the form 
$\int dt(p\dot q-H(p,q))$, where $H$ is the Hamiltonian and  
$t$, the coordinate in the bulk direction,
plays the role of the time. For the quadratic action 
this prescription implies that it does not contain
any gravity or scalar fields with 
second derivative in the bulk direction.  
All such terms are integrated by parts and arising 
boundary terms are simply discarded.    

{}For the sake of simplicity we consider the case of a unique 
active scalar $\bar{\phi}$ and pick up the ``almost'' radial gauge
for which the scalar fluctuation $\phi$ is set to zero. 
The residual gauge symmetry is enough to decouple the  
transverse components of the graviton and it leads to the 
conservation law for the stress-energy tensor in the boundary
field theory. Except for the traceless transverse graviton 
the only physical degree of freedom is then 
the trace of the graviton $h$ 
for which we obtain a very simple action,  basically 
due to the existence of a superpotential.  

The trace of the graviton is usually viewed as the gravity 
field dual to the trace of the stress-energy tensor in the field theory 
away from the conformal point. However, we show that by using a field 
redefinition one can recast the action for $h$ into the standard action 
for a scalar field $s$ in the kink background with some complicated 
potential. Depending of the form of the kink solution, 
the scalar $s$ may then be naturally interpreted 
as the dual either to the YM operator invoking RG flow or to 
the corresponding field theory operator with 
a non-trivial vacuum expectation value. 
The relation between $h$ and $s$ appears to be $h\sim \b^{-1} s$, where 
$\b $ is the holographic beta function 
introduced in \cite{AGPZ}. We believe that a similar relation also 
occurs for general backgrounds with many active scalars 
\cite{aft2}.

We then consider the same two kink solutions as in \cite{DF}.
The first one corresponds to ${\cal N}=4$ SYM,
perturbed by an operator of dimension 3, 
that flows in the IR to ${\cal N}=1$ SYM.
The second kink involves one active scalar from the ${\bf 20}$ of $SO(6)$
and the corresponding dual flow describes the states on the Coulomb branch 
of ${\cal N}=4$ SYM, which is parametrized by the vacuum 
expectation value of one of the operators $O_2=\tr (X^{(I}X^{J)})$, where 
$X^I$ are YM scalars.

Evaluating the on-shell action for these supergravity 
solutions we find that in both cases the two-point functions 
exhibit the same behaviour as was found for the inert 
scalars in \cite{FGPW2,DF}.
Indeed, in the first case we get a discrete spectrum 
while in the second example the spectrum is continuous with a mass gap. 

\section{Gravity/active scalar system}
Consider $(d+1)$-dimensional gravity,
described by a metric $G_{\mu\nu}$ with signature $(1,1,...,1,-1)$,
minimally coupled to scalar fields $\varphi^I$. 
The action is of the form
\footnote{Our conventions are: $[\n_\mu ,\n_\nu ]V_\rho 
= R_{\mu\nu\rho}{}^\s V_\s$
and $R_{\mu\rho}=R_{\mu\nu\rho}{}^\nu $. }
\bea
S = \int\, d^{d+1}x \sqrt{-G} \l \frac14 R - 
\frac12 G^{\mu\nu}\p_\mu\varphi^I\p_\nu\varphi^I - V(\varphi )\r ;
\la{action}
\eea
$V(\varphi )$ is the scalar potential.
The equations of motion that follow from this action are 
\bea
&&R_{\mu\nu} -\frac12 G_{\mu\nu}R-2\p_\mu\varphi^I\p_\nu\varphi^I
+G_{\mu\nu}\p_\rho\varphi^I\p^\rho\varphi^I +2G_{\mu\nu}V=0,\cr
&&\n^\mu\p_\mu\varphi^I - \frac{\p V}{\p\varphi^I} =0.
\la{eqs1}
\eea
They imply
\bea
&&R-2\p_\rho\varphi^I\p^\rho\varphi^I = 4\frac{d+1}{d-1}V\,,\cr
&&R_{\mu\nu} -2\p_\mu\varphi^I\p_\nu\varphi^I = \frac{4V}{d-1}G_{\mu\nu}\,.
\la{eqs2}
\eea
Let $g_{\mu\nu}$ and $\bar{\phi}^I$ be solutions of the equations of
motion and decompose $G_{\mu\nu}$ and $\varphi$ around 
their background values
$$
G_{\mu\nu} = g_{\mu\nu} +h_{\mu\nu},\qquad 
\varphi^I =  \bar\phi^I+\phi^I .
$$
Discarding total derivative terms,  
we find the following quadratic action for the fluctuations
\bea
S_2 = \int\, d^{d+1}x \sqrt{-g} &\biggl[& \frac14\biggl( 
- \frac14\n_\rho h_{\mu\nu}\n^\rho h^{\mu\nu} + 
\frac12\n_\rho h_{\mu\nu}\n^\mu h^{\rho\nu} \cr
&-&
\frac12\n^\nu h\n^\mu h_{\mu\nu}+
 \frac14\n_\mu h\n^\mu h +
\frac{2V}{d-1}h^2_{\mu\nu}-
\frac{V}{d-1}h^2 \biggr)\cr
&+& h^{\mu\nu}\n_\mu\phi^I\n_\nu\bar{\phi}^I+
\frac12 \n_\mu h\phi^I\n^\mu\bar{\phi}^I \cr
&-& 
\frac12\n_\mu\phi^I\n^\mu\phi^I - \frac12 V_{IJ}\phi^I\phi^J\biggr] .
\la{action2}
\eea
The covariant derivative $\n_\mu$ is with respect to 
the background metric $g_{\mu\nu}$ which is also used to raise and 
lower indices, $h=h_\mu^\mu$, and we have introduced the notation
$$V_I=\frac{\p V}{\p\varphi^I}|_{\varphi =\bar{\phi}},\qquad
V_{IJ}=\frac{\p^2 V}{\p\varphi^I\p\varphi^J}|_{\varphi =\bar{\phi}}.
$$
Within the context of the AdS/CFT correspondence, 
we are interested in background solutions
which respect $d$-dimensional Poincar\'e invariance. Thus,
we make an ansatz for the background  
solving (\ref{eqs1}) of the form
\bea
ds^2 = dx_0^2 + e^{2A(x_0)}\eta_{ij}dx^idx^j,
\la{metric}
\eea
where $\eta_{ij}$ is the Minkowski metric and take $\bar{\phi}$ 
to depend only on $x_0$. 

The analysis of the action and the equations of motion is complicated by
the fact that the scalar fields couple to the graviton 
already in the quadratic action. 
Therefore, the spectrum of the theory on this background
cannot be readily read off the action or the equations. 
However, the problem of finding the spectrum of the theory can be
simplified by an appropriate gauge choice. 
The quadratic action and the equations of motion are invariant
under gauge transformations induced by reparametrizations:
\bea
\d h_{\mu\nu} = \n_\mu\zeta_\nu +\n_\nu\zeta_\mu, \quad 
\d\phi^I =\zeta^\mu\n_\mu \bar{\phi}^I .
\la{inv}
\eea
One usually imposes the temporal gauge $h_{0\mu}=0$ and solves 
the constraints, i.e. the equations of motion for $h_{0\mu}$.   
This was done in \cite{DF}.
However, this gauge choice leads to a very complicated 
system of equations and, moreover, it was only possible to derive
a third-order equation for scalar fields. On the other hand,
the gauge transformations (\ref{inv}) of scalar fields show that on  
backgrounds with at least one nonvanishing field 
$\bar{\phi}^1\equiv\bar{\phi}$
one can impose the almost temporal gauge 
$h_{0i}=0=\phi$. This gauge choice is very natural, because 
the combination 
$\tilde{h}_{00}\equiv h_{00} - 2\n_0 (\phi /\n_0 \bar{\phi})$
is gauge invariant, and one could express the action and the 
equations of motion in terms of $\tilde{h}_{00}$ rather than $h_{00}$.
Then only the scalar fields change under gauge transformations. 

{}For the sake of simplicity we 
restrict ourselves to considering 
the simplest case of one active scalar.  
Then the action depends only on $h_{ij}$ and $h_{00}$.
It is convenient to introduce $t_{ij}$ via 
$$h_{ij}=e^{2A}t_{ij}, \quad h^{ij}=e^{-2A}t^{ij},
\quad t\equiv t_i^i=t_{ij}\eta^{ij}\,.$$
In what follows we will not distinguish upper and lower indices.
It is straightforward to derive
\bea
&&\n_0h_{00}=\p_0h_{00},\quad \n_ih_{00}=\p_ih_{00},\quad 
\n_0h_{0i}=0,\cr
\noalign{\vskip.2cm}
&&\n_ih_{0j}=e^{2A}\p_0A(\eta_{ij}h_{00}-t_{ij}),\quad 
\n_0h_{ij}=e^{2A}\p_0t_{ij},\quad 
\n_ih_{kl}=e^{2A}\p_it_{kl}
\nonumber
\eea
and to cast the action (\ref{action2}) in the form
\bea
S_2 &=& \int\, d^{d+1}x \, \frac14 e^{dA}
\biggl[\, - \frac14 (\p_0t_{ij})^2 + \frac14 (\p_0t)^2- 
 \frac14 e^{-2A}(\p_it_{kl})^2 + \frac12 \p_0^2At_{ij}^2\cr
&+& \frac{d}{2}(\p_0A)^2t_{ij}^2+\frac12 e^{-2A}\p_it_{kl}\p_kt_{il}-
\frac14 \p_0^2At^2-\frac{d}{4}(\p_0A)^2t^2\cr
&+& \frac12 e^{-2A}t\p_i\p_jt_{ij}+\frac14e^{-2A}(\p_it)^2+\frac{2V}{d-1}t_{ij}^2
-\frac{V}{d-1}t^2\cr
&+&h_{00}\biggl(\, \frac{1-d}{2}\p_0A\p_0t -\frac12 \p_0^2At
-\frac{d}{2}(\p_0A)^2t+\frac12 e^{-2A}\p_i\p_jt_{ij}-\frac12e^{-2A}\p_i^2t-
\frac{2V}{d-1}t\,\biggr) \cr
&+& h_{00}\biggl(\, \frac{d}{4}\p_0^2A + \frac{d^2}{4}(\p_0A)^2
+\frac{V}{d-1} \,\biggr) h_{00}\,\biggr]\,.
\la{act2}
\eea
We see from this action that $h_{00}$ is a non-dynamical field, 
and can thus be integrated out.
On the other hand we have the constraints that follow from the
$h_{0i}$ equations of motion and 
from the equation for $\phi$. They are
\bea
&&(1-d)\p_0A\p_ih_{00} + \p_0(\p_it-\p_jt_{ji}) =0\,,
\la{con}\\
&&\p_0h_{00}\p_0\bar{\phi}+2(d\p_0A\p_0\bar{\phi}
+\p_0^2\bar{\phi})h_{00}-
\p_0t\p_0\bar{\phi}=0.
\la{con2}
\eea
These constraints allow us to express $h_{00}$ through $t_{ij}$ 
and therefore, they should be compatible with the equation of motion
for $h_{00}$ that follows from (\ref{act2}). 

{}From here on we restrict ourselves to the most interesting case $d=4$, and 
to a potential $V(\varphi )$ which can be derived from a 
superpotential $W(\varphi )$,
\be
V(\phi ) = \frac{g^2}{8} \l \frac{\p W}{\p\phi}\r^2 - \frac{g^2}{3} W^2.
\la{sp}
\ee
All explicitly known backgrounds are obtained from such a potential.
One can show \cite{FGPW1}, that any solution to the 
equations
\be
\p_0 A = -\frac{g}{3} W,\quad \p_0\bar{\phi} 
=\frac{g}{2} \frac{\p W}{\p\bar{\phi}},
\quad g=\frac{2}{L},
\la{rel}
\ee
also satisfies the equations of motion (\ref{eqs1}).
The length scale $L$ is related to the cosmological 
constant $\Lambda$ via $\Lambda = -12/L^2 = 4V(\varphi =0)$.
It is not difficult to verify that these 
relations lead to the identity
\bea
\p_0^2 A+4(\p_0A)^2+\frac43 V =0.
\la{id}
\eea
This identity simplifies the action  (\ref{act2}) considerably and 
it now takes the form
\bea
S_2 &=& \int\, d^5x \, \frac14 e^{4A}
\biggl[\, - \frac14 (\p_0t_{ij})^2 + \frac14 (\p_0t)^2- 
 \frac14 e^{-2A}(\p_it_{kl})^2 
+\frac12 e^{-2A}\p_kt_{kl}\p_it_{il}\cr
&+& \frac12 e^{-2A}t\p_i\p_jt_{ij}+\frac14e^{-2A}(\p_it)^2\cr
&+&h_{00}\biggl(\, -\frac{3}{2}\p_0A\p_0t
+\frac12 e^{-2A}\p_i\p_jt_{ij}-\frac12e^{-2A}\p_i^2t\,\biggr) 
-Vh_{00}^2\,\biggr]\,.
\la{act24}
\eea
It is well-known (see, e.g. \cite{AF1}) that the transverse 
traceless components of the metric fluctuations decouple,
and are described by the same equation as a free minimally-coupled 
massless scalar in the background (\ref{metric}).
The physical reason for the decoupling is that
due to the Lorentz invariance the boundary stress tensor is conserved
and, therefore, only the transverse traceless 
components can couple to it.
{}For this reason it is convenient to introduce 
the following decomposition of the graviton
$$
t_{ij}=t_{ij}^{\perp}+t_{ij}^{||}+\frac{1}{4}h\eta_{ij}-\p_i\p_j H.
$$
Here $t_{ij}^{\perp}$ is the traceless transverse part 
and $t^{||}_{ij}$
is a traceless longitudinal part given by 
\bea
t_{ij}^{||}=\frac{\p_i}{\Box} \p_kt_{jk}+\frac{\p_j}{\Box} \p_kt_{ik}
-2\frac{\p_i\p_j}{\Box^2} \p_k\p_mt_{km} ; 
\eea
it satisfies $\p_i\p_jt_{ij}^{||}=0$. 

Substituting this decomposition in the action one can easily
see that the  transverse traceless components do
decouple, and in what follows we will drop them. 
Moreover, the longitudinal traceless components also decouple and 
the only remaining coupled fields are $h$ and $H$.

To analyze their action  
we introduce their Fourier transforms 
$$t_{ij}(x_0, x) =\frac{1}{4\pi^2}\int\,d^4p\, e^{ipx}t_{ij}(x_0,p),
\quad h_{00}(x_0, x) =\frac{1}{4\pi^2}\int\,d^4p\, e^{ipx}h_{00}(x_0,p).
$$
In momentum space the constraints (\ref{con}) take the form
\bea
-3\p_0A h_{00}p_i+\frac{3}{4} p_i\p_0 h - p_j\p_0 t_{ij}^{||}=0.
\nonumber
\eea 
{}From here one finds that 
\be
\p_0Ah_{00}=\frac14\p_0h.
\la{0i}
\ee
and $t_{ij}^{||}$ does not depend on $x_0$. 
Thus,  $t_{ij}^{||}$ are not dynamical modes and do not couple 
to any operators in the boundary theory. They may therefore be omitted.
 
We are left with the graviton modes $t_{ij}$ of the form
$$t_{ij}=\frac14\eta_{ij}h+p_ip_jH,$$
whose dynamics is described by the action
\bea
S_2 &=& \int\, dx_0d^4p \, \frac14 e^{4A}
\biggl[\, \frac{3}{16}(\p_0h)^2 + \frac{3}{8}p^2\p_0h\p_0H +
\frac{3}{32}e^{-2A}p^2h^2\cr
&+&
h_{00}\biggl(\, -\frac32\p_0A(\p_0h+p^2\p_0H)+
\frac38 e^{-2A}p^2h\,\biggr) - Vh_{00}^2\,\biggr]\,. 
\la{act4}
\eea
Making the following shift of $h_{00}$ 
\be
h_{00}\to h_{00} + \frac{\p_0h}{4\p_0A},
\la{shift}
\ee
we rewrite (\ref{act4}) as
\bea
S_2 &=& \int\, dx_0d^4p \, \frac14 e^{4A}
\biggl[\, \frac{3}{16}(\p_0h)^2 -\frac{3}{2}p^2\p_0Ah_{00}\p_0H +
\frac{3}{32}e^{-2A}p^2h^2\cr
&+&
\l h_{00}+ \frac{\p_0h}{4\p_0A}\r 
\biggl(\, -\frac32\p_0A\p_0h+
\frac38 e^{-2A}p^2h\,\biggr) - V\l h_{00}
+ \frac{\p_0h}{4\p_0A}\r^2\,\biggr]\,. 
\la{act5}
\eea
Thus $h_{00}$ is a momentum for $H$ and the constraint
(\ref{0i}), which, after the shift (\ref{shift}) reads $h_{00}=0$,  
shows that $H$ is not a dynamical field. 
So we can set $h_{00}=0$
and obtain, using again (\ref{id}), the final action for $h$
\bea
S_2 =\int\, dx_0d^4p \, \frac14 e^{4A}\cdot
 \frac{3}{64} \frac{\p_0^2A}{(\p_0A)^2}\biggl(\,
(\p_0h)^2 +e^{-2A}p^2h^2
\biggr)\,.
\la{actf}
\eea
Some comments are in order. The action (\ref{actf})
has the correct overall sign 
as $\p_0^2A$ is always negative \cite{FGPW1}. 
The terms in parentheses are exactly
the same as for the transverse traceless components. Actually
the only difference between the action (\ref{actf}) and the action
for the transverse traceless components is in the factor
$$-\frac{3}{16}\frac{\p_0^2A}{(\p_0A)^2}=
\frac{9}{32}\biggl(\,\frac{\p}{\p\bar{\phi}}{\rm log} W\,\biggl)^2 
=\frac18 \l \frac{d\bar{\phi}}{dA}\r^2=\frac18 \beta^2,
$$
where $\b = \frac{d\bar{\phi}}{dA}$ is the holographic beta function 
introduced in \cite{AGPZ}. 
One can remove this factor by rescaling $h$
$$h=\frac{8}{\b}s=8\sqrt{-\frac{2(\p_0A)^2}{3\p_0^2A}}s.$$
The new field $s$ is a scalar with proper 
transformation properties under reparametrizations, however,
it has a complicated potential $U(x_0)$, and is described by the action
\bea
S_2 =-\int\, dx_0d^4p \, \frac12 e^{4A}\,\biggl(\,
(\p_0s)^2 +(e^{-2A}p^2 +U(x_0))s^2\biggr),
\la{acts}
\eea
where
\bea
\la{pot}
U(x_0)=2\l\frac{A''}{A'} \r^2-2\frac{A'''}{A'}
-\frac{1}{4}\l\frac{A'''}{A''} \r^2
+\frac{1}{2}\frac{A^{(iv)}}{A''}+2\frac{A'''A'}{A''}-4A''.
\eea

The appearance of the relative factor between $h$ and $s$ 
is indeed very natural, because the trace of
the graviton $h$ is dual to the trace of the stress tensor and,
as was discussed in \cite{DF}, in a deformed conformal 
field theory we expect to have an operator relation
\cite{Osborn}
$$T_i^i(x) = \b{\cal O}(x),$$
where ${\cal O}$ is the operator responsible for the deformation of
the conformal field theory. 
In other words, the coupling of $h$ to the trace of the stress-energy
tensor is $\int_\e h T=\int_e {1\over\beta}sT\equiv 
\int_\e s{\cal O}$.
Therefore, we see that if $h$ is dual to the trace of the stress tensor
then the scalar $s$ is dual to the operator ${\cal O}$.

It is worth noting that we have no linear term in the final action.
Linear terms do not appear
because they can come only from total-derivative bulk
terms, but we omit any such a term following the prescription of \cite{AF1}.
Moreover, according to \cite{AF1}, we do not need to add any boundary
term to the actions (\ref{actf}) or (\ref{acts}), and these actions are 
appropriate for computing the 2-point function of the operator $\cal{O}$.
The absence of linear terms leads to the vanishing of 
the one-point correlator of the operator $\cal{O}$ dual to the scalar $s$.
This may look strange because one usually says that in a perturbed
conformal field theory an operator dual to an active scalar has a 
nonvanishing one-point function.  
Nevertheless, we work in flat 4-dimensional space, and, therefore, 
we can always choose such a substraction scheme 
that a one-point function of any local operator vanishes.  
This seems to mean that the scalar $s$ is actually dual to the operator
${\cal O} - \langle {\cal O}\rangle$.

Deriving the action (\ref{actf}), we have not used the constraint
(\ref{con2}) yet. One may wonder if this constraint imposes additional
restrictions on admissible configurations of the gravity fields.
In the appendix we show that this constraint follows from the 
equations of motion for $h$ and $H$, and from the $h_{0i}$-constraints
(\ref{con}), and, therefore, can be omitted.
  
Next we use the action (\ref{actf}) to compute 2-point functions
in the two cases recently studied in \cite{DF}.
Recall that the 2-point functions of active scalars obtained in \cite{DF}
appear to be problematic. We will see 
that action (\ref{actf}) in both cases leads to 
reasonable 2-point functions.

We begin with the case considered in section 3 of \cite{DF}. The 
supergravity solution discussed there was found in \cite{GPPZ},
and describes the renormalization group flow of ${\cal N}=4$ SYM theory 
to ${\cal N}=1$ SYM theory at long distances. 
We refer the reader to \cite{GPPZ} for details.

To simplify the equations of motion 
of the scalar field and its solution,  
a new coordinate $u$ was introduced in \cite{DF} such that
$$e^{2A}=\frac{u}{1-u},\quad \frac{du}{dx_0}=\frac{2}{L}(1-u),\quad 
\frac{dA}{dx_0}=\frac{1}{Lu}.$$
In this coordinate the boundary of the 5-d space is at $u=1$, and 
there is a singularity at $u=0$.
With these formulas we rewrite the action (\ref{actf}) as 
\bea
S_2 =-\frac{3}{64L}\int\, du\,d^4p \, u^2\biggl(\,
(\p_uh)^2 +\frac{p^2L^2}{4u(1-u)}h^2
\biggr)
=-{3\over 64L}\int\, du\,d^4p\,\partial_u(u^2h\partial_u h)\,,
\la{actfu}
\eea
where in the second step the equation of motion for $h$ 
\be
\p_u^2h +\frac{2}{u}\p_uh-\frac{p^2L^2}{4u(1-u)}h =0
\la{eq1u}
\ee
has been used. 
To compute the 2-point function we follow the standard AdS/CFT 
prescription \cite{GKP,W}. Imposing Dirichlet boundary conditions 
at $u=1-\e^2$, we obtain the solution to (\ref{eq1u}), regular at $u=0$
\be
h(u,p)=\frac{F(a_-,a_+;2;u)}{F(a_-,a_+;2;1-\e^2)}h(p)\,,
\la{solu}
\ee
where
$$a_\pm =\frac12 (1\pm\sqrt{1-p^2L^2})$$
and $F(a,b;c,u)$ is the hypergeometric function ${}_2F_1$.
The 2-point function of $h$ in momentum space 
is given by the familiar formula
\bea
\langle O(p)O(-p)\rangle ={\rm lim}_{\e\to 0}\frac{3}{32L}
 \frac{\p_u F(a_-,a_+;2;u)}{F(a_-,a_+;2;u)}|_{u=1-\e^2}\,.
\la{corru}
\eea
To find the 2-point function we need 
$$\frac{d}{du}(uF(a,b;2;u))=F(a,b;1;u),\quad 
F(a_-,a_+;2;1)=\frac{1}{\G (2-a_-)\G (2-a_+)},$$
and the expansion
$$F(a_-,a_+;1;u) = \frac{1}{\G (a_-)\G (a_+)}\l 2\Psi (1) -\Psi (a_-)-
\Psi (a_+) - 2{\rm log}\e + o(\e )\r .$$
Then, omitting all terms polynomial in $p$, we find in the limit $\e\to 0$
\bea
\langle O(p)O(-p)\rangle =\frac{3}{32L}
 \frac{p^2L^2}{4}\l \Psi (\frac12 (1-\sqrt{1-p^2L^2}))+
\Psi (\frac12 (1+\sqrt{1-p^2L^2})) \r\,.
\la{corru1}
\eea
The correlator has a discrete spectrum of poles at $-p^2=4n(n+1)/L^2$,
$n=0,1,...$. A similar discrete spectrum was found in \cite{GPPZ,DF} 
by studying 2-point correlators of inert scalars. Thus contrary to
the result in \cite{DF}, we obtain for the active scalar
a 2-point correlator with the expected behaviour.

Next we consider the second case, c.f. section 4 of \cite{DF}, where
the flow obtained in \cite{FGPW2,BS} was used to analyze the 2-point 
function of an active scalar. This flow changes the vacuum of 
${\cal N}=4$ SYM which is now on the Coulomb branch. 

A new coordinate $v$ was introduced \cite{DF} such that
\bea
e^{2A}=\frac{l^2}{L^2}\frac{v^{2/3}}{1-v},
\quad \frac{dv}{dx_0}=\frac{2}{L}v^{2/3}(1-v),\quad 
\frac{dA}{dx_0}=\frac{v+2}{3Lv^{1/3}}\,,
\la{kink2}
\eea
$l$ is an additional length scale. 
As in the first example, 
the boundary of the 5-d space is at $v=1$, and 
there is a singularity at $v=0$.

{}First we check that our interpretation of the field $s$ as the dual
to the operator $O_2$ in the YM is consistent.
{}For the kink solution (\ref{kink2}) there exists a
limiting procedure \cite{FGPW2}
that allows one to remove the flow and to restore the AdS solution.
This is a limiting case when  
$\frac{l^2}{L^2}\equiv\xi^2$ becomes small 
while $\frac{z}{L}=x$ is kept fixed.
The variable $z$ is introduced via
$$
v={\rm sech}^2\l \frac{zl}{L^2} \r=\frac{1}{{\rm ch}^2(\xi x)}
=1-\xi^2x^2+\frac{2}{3}\xi^4x^4+...
$$ 
Clearly, in this limit 
$$
2A=\log\l \xi^2\frac{v^{2/3}}{1-v} \r \approx -\log(x^2),
$$
and from $\frac{dv}{dx_0}=\frac{2}{L}v^{2/3}(1-v)$
it follows that 
\bea
\la{dr}
dx_0=-\frac{L}{x}dx ,
\eea
i.e. one recovers the standard AdS metric. 

By using $A'=dA/dx_0$ from (\ref{kink2}) together with (\ref{dr}) 
it is  easy to find 
the leading terms of the derivatives 
\be
A'\sim\frac{1}{L}+\frac{1}{9L}\xi^4x^4;~~~
A''\sim-\frac{4}{9L^2}\xi^4x^4;~~~
A'''\sim\frac{4^2}{9L^3}\xi^4x^4;~~~
A^{(IV)}\sim-\frac{4^3}{9L^4}\xi^4x^4;
\la{approx}
\ee
Since $\beta\sim\xi^2$, it vanishes in this limit.  

Upon substituting (\ref{approx}) into (\ref{pot}) one gets 
$U(x_0)=-\frac{4}{L^2}+{\cal O}(\xi^4)$. Thus, in the limit $\xi\to 0$
the action (\ref{acts}) becomes
\bea
S_2=-\frac{L^3}{2}\int dz d^4p \sqrt{-g_a}\l z^2(\p_z s)^2+
z^2p^2 s^2- 4s^2 \r , 
\la{stac}
\eea
where $g_a$ is the determinant of the standard $AdS$ metric. Eq. (\ref{stac})
is the familiar action for the scalar field on the AdS space with mass 
$m^2=-4$ that is dual to the YM operator $O_2$ of conformal weight $\D=2$.
It is worth noting that from the point of view of the 2-point correlator of
the operator dual to the scalar $s$, the limit $\xi\to 0$ is equivalent to
taking $p^2$ to infinity, i.e. to the UV limit. 
Therefore, this consideration shows that 
the 2-point correlator does not vanish, and behaves itself in
the UV as expected from an operator of the UV conformal dimension $\D =2$.

With the help of (\ref{kink2}) we rewrite (\ref{acts}) as 
\bea
S_2 =-\frac{l^4}{L^5}\int\, dvd^4p \, \frac{v^2}{(1-v)}
\biggl(\,
(\p_v s)^2 +\frac{p^2L^4}{4l^2v^2(1-v)}s^2
-\frac{3(4-2v+v^2)}{(v+2)^2v(1-v)^2}s^2\,.
\biggr)
\la{actfv}
\eea
This action leads to the following equation of motion for $s$
\be
\p_v^2s +\frac{2-v}{v(1-v)}\p_vs-\frac{p^2L^4}{4l^2v^2(1-v)}s+
\frac{3(4-2v+v^2)}{(v+2)^2v(1-v)^2}s
 =0\,.
\la{eq1v}
\ee
The on-shell value of the action is 
\bea
S_2 =-\frac{l^4}{L^5}\int\, 
d^4p \, \l \frac{v^2}{1-v}s\p_v s \r_{v=1-\eps^2} . 
\la{onsh}
\eea 
Equation (\ref{eq1v}) 
has four regular singular points, $(-2,0,1,\infty)$ and a closed form 
solution does not exist. But we can nevertheless analyze the 
behaviour of a solution in the neighbourhood
of the physically
relevant points at $v=1$ and at $v=0$. At $v=1$, the 
power series Ansatz $s(v)=(1-v)^\rho[1+a_1(1-v)+{\cal O}((1-v)^2)]$ 
leads to the indicial  
equation $\rho^2-2\rho+1=0$ with two degenerate solutions
$\rho=1$.
Thus, the general solution is of the form 
\be
s(v)=(1-v)\l c_1 f_1(v) + c_2f_2(v) \log (1-v)\r ,
\la{s(v)}
\ee
where $f_1$ and $f_2$ are power series in $(1-v)$ 
with leading term 1. 

A similar analysis at $v=0$ leads to an indicial equation with the 
two solutions
$$
\rho_{\pm}=-\frac{1}{2}\pm \frac{1}{2}\sqrt{1+\frac{p^2L^4}{l^2}}.
$$ 
{}For generic parameters the two independent solutions are pure power series.
Of these, the one for $\rho_-$, is forbidden by regularity. 
Thus, the solution we are interested in has the general form 
$$
h(v)=c v^{\rho_+}(1+ a_1 v+a_2 v ^2+\dots)\,.
$$
It is regular at $v=0$ for space-like momenta ($p^2>0$).
Note however that although the solution is not regular at 
the singularity 
$v=0$ if the momentum obeys $0\le-p^2< l^2/L^4$,  
the $v$-integration, c.f. (\ref{onsh}), gives a vanishing 
contribution at $v=0$ due to the factor $v^2$ in the numerator.

To find the 2-point function we need to analytically 
continue this solution to the neighbourhood of $v=1$. 
By comparing with the known solution of (\ref{eq1v}) 
in the UV limit 
$p^2\to\infty$ ($\xi\to 0$) we conclude that 
both constants $c_1$ and $c_2$ in (\ref{s(v)}) are nonvanishing. 
It is not difficult to compute the 2-point function in terms of 
these constants. By using the conventional rules of the AdS/CFT 
correspondence, we get
\bea
\langle O(p)O(-p)\rangle ={\rm lim}_{\e\to 0}\frac{2l^4}{L^5}
 \frac{v^2}{1-v}\p_vs_\e (v)|_{v=1-\e^2},
\la{corrv}
\eea
where 
$$
s_\e (v) =\frac{(1-v)\l c_1 f_1(v) + c_2f_2(v) \log (1-v)\r}
{\e^2\l c_1 f_1(1-\e^2 ) + c_2f_2(1-\e^2) \log (\e^2)\r}
$$
is the solution of the equation of motion normalized to be 1 at $v=1-\e^2$.
The leading term in $\e$, non-analytic in $p^2$, is
\bea
\langle O(p)O(-p)\rangle =\frac{1}{\e^4\log^2\e}\cdot\frac{l^4c_1}{2L^5c_2}.
\la{corrvf}
\eea
The factor $\frac{1}{\e^4\log^2\e}$ is the one that one expects  
in the 2-point correlator of a scalar field with the UV conformal weight 2
(see, e.g. \cite{MR} or appendix of \cite{AF2}). 
Although 
we do not know the constants $c_1$ and $c_2$ explicitly, their
$p^2$-dependence will be through $\rho_+$. We thus conclude, 
as in \cite{DF} for the case of the inert scalar,  
that the correlator has a continuous spectrum with a mass gap, 
$m^2\geq l^2/L^4$, which vanishes in the limit $\xi\to 0$.
\section{Conclusion}
In this paper we studied graviton-scalar fluctuations 
in $d=5$ flow geometries which are dual to boundary field theories
with RG flow. We showed that the analysis of the coupled gravity-scalar 
sector drastically simplifies by a choice of the almost radial gauge,
where one of the active scalars vanishes, but the trace of the 
graviton $h$ remains dynamical. We considered in detail 
the simplest case of one active scalar, decoupled the graviton trace
from the transverse traceless components, and obtained a very simple 
quadratic action for it. The Lagrangian of the graviton trace
differs from the one of a minimally-coupled massless scalar field 
(e.g. dilaton) 
only by a factor which coincides with the square of the 
holographic beta function of the operator responsible for the
deformation of the conformal field theory. This is a very natural result
due to the operator relation $T_i^i=\sum_I\b^I{\cal O}^I$ 
in a deformed CFT. We expect that a similar relation between 
the action for $h$ and the dilaton action also holds
for the general case of many active scalars, where the factor 
would be given by the square of 
a ``weighted'' holographic beta function.   

We fixed the form of the quadratic action by means of the 
Hamiltonian prescription \cite{AF1} and
used the action to compute 2-point functions of 
the graviton trace in two cases of flow geometries.
In both cases we obtained physically reasonable functions
which have the same momentum dependence as those of inert scalars.
Thus, we successfully resolved the problem recently 
raised in \cite{DF}. 

It is worth noting that the geometries we considered have 
a curvature singularity in the interior and one would expect 
large string corrections to the 2-point functions, which could
drastically change the behaviour of the correlators. Nevertherless,
this seems not to happen because the curvature is small 
near the boundary, and the contribution of 
the vicinity of the singularity to 
the on-shell gravity action vanishes.

\appendix{The second constraint}
Here we analyze the constraint (\ref{con2}) and show 
that it follows from (\ref{con}) and the equations of motion. 
Since the equation of motion for $\bar{\phi}$ is 
$$
\frac{\p V}{\p\bar{\phi} }=\p_0^2\bar{\phi}+4\p_0\bar{\phi}\p_0 A\,,
$$
eq.(\ref{con2}) can be written as  
\bea 
C=\p_0h_{00}\p_0\bar{\phi}+2\frac{\p V}{\p\bar{\phi} }h_{00}-
\p_0t\p_0\bar{\phi}.
\eea
By using the equation of motion for $h_{00}$ together
with (\ref{con}) we find
$$
p^2\p_0H=-\p_0h-\frac{V}{3}\frac{\p_0h}{(\p_0A)^2}
+\frac{1}{4}\frac{e^{-2A}p^2h}{\p_0 A}
$$
and, therefore, 
$$
\p_0t= -\frac{V}{3}\frac{\p_0h}{(\p_0A)^2}
+\frac{1}{4}\frac{e^{-2A}p^2h}{\p_0 A}.
$$
Taking into account that 
\bea
\nonumber
\p_0h_{00}=
\frac{1}{4}\frac{\p_0^2h}{\p_0A}
-\frac{1}{4}\frac{\p_0^2A}{(\p_0A)^2}\p_0h
\eea 
one finds 
\bea
C=\l \frac{1}{4}\frac{\p_0^2h}{\p_0A}
-\frac{1}{4}\frac{\p_0^2A}{(\p_0A)^2}\p_0h
-\frac{1}{4}\frac{e^{-2A}p^2h}{\p_0 A}
+\frac{V}{3}\frac{\p_0h}{(\p_0A)^2}\r \p_0\bar{\phi}
+\frac{1}{2}\frac{\p V}{\p\bar{\phi} }\frac{\p_0 h}{\p_0 A}
\eea
{}From (\ref{actf}) one finds the equation of motion for $h$:
\bea
\frac{\p_0^2 h}{\p_0A}-2\frac{\p_0^2A}{(\p_0A)^2}\p_0h
-\frac{e^{-2A}p^2h}{\p_0 A}=-4\p_0h-\frac{\p_0^3A}{\p_0A\p_0^2A}\p_0h.
\eea  
Therefore, by virtue of this equation one obtains
\bea
C=\l \frac{1}{4}\frac{\p_0^2A}{(\p_0A)^2}\p_0h
-\p_0h-\frac{\p_0^3A}{4\p_0A\p_0^2A}\p_0h
+\frac{V}{3}\frac{\p_0h}{(\p_0A)^2}\r \p_0\bar{\phi}
+\frac{1}{2}\frac{\p V}{\p\bar{\phi} }\frac{\p_0 h}{\p_0 A}.
\eea
Recalling (\ref{id}), one then finds
\bea
C= \frac{1}{4}\l 
-8-\frac{\p_0^3A}{\p_0A\p_0^2A}
\r \p_0\bar{\phi}\p_0h
+\frac{1}{2}\frac{\p V}{\p\bar{\phi} }\frac{\p_0 h}{\p_0 A}.
\eea
Differentiating (\ref{id}), one gets 
$$
\p_0^3A+8\p_0A\p_0^2A+\frac{4}{3}\frac{\p V}{\p\bar{\phi} }\p_0\bar{\phi}=0
$$
and, as a consequence,
$$
C=\frac{1}{3}\frac{\p V}{\p\bar{\phi} }
\frac{(\p_0\bar{\phi})^2\p_0 h }{\p_0 A\p_0^2A} 
+\frac{1}{2}\frac{\p V}{\p\bar{\phi} }\frac{\p_0 h}{\p_0 A}=
\frac{1}{2}\frac{\p V}{\p\bar{\phi} }\frac{\p_0 h}{\p_0 A\p_0^2A}
\l \frac{2}{3}(\p_0\bar{\phi})^2+ \p_0^2A  \r .
$$
{}Finally, we have the relation
\bea
\p_0^2A=-\frac{g}{3}\frac{\p W}{\p \phi}\p_0\bar{\phi}=
-\frac{2}{3}(\p_0\bar{\phi})^2.
\eea
Upon substituting this relation into the previous formula we find
$C=0$. Thus, constraint (\ref{con2}) is compatible with the dynamics.

\vskip 1cm
{\bf ACKNOWLEDGMENT}
We would like to thank A. Tseytlin for useful comments.
The work of G.A. was
supported by the Alexander von Humboldt Foundation and in part by the
RFBI grant N99-01-00166, and the work of S.F. was supported by
the U.S. Department of Energy under grant No. DE-FG02-96ER40967 and
in part by RFBI grant N99-01-00190. S.T. is supported by GIF -- 
the German-Israeli Foundation for Scientific Research
by the TMR programme ERBRMX-CT96-0045.


\begin{thebibliography}{99}
{\small
\bibitem{M} J. Maldacena, 
Adv. Theor. Math. Phys. 2 (1998) 231-252.
\bibitem{GKP} G.G. Gubser, I.R. Klebanov and A.M. Polyakov,  
Phys.Lett. B428 (1998) 105-114, hep-th/9802109.
\bibitem{W} E. Witten,  
Adv.Theor.Math.Phys. 2 (1998) 253-291, hep-th/9802150.
\bibitem{GPPZ0} L. Girardello, M. Petrini, M. Porrati, and
A. Zaffaroni, JHEP 12 (1998) 022, hep-th/9809047.
\bibitem{DZ} J. Distler and F.Zamora, Adv. Theor. Math. Phys. 2 (1999) 1405,
hep-th/9810206. 
\bibitem{KPW} A. Khavaev, K. Pilch and N. P. Warner, New vacua of gauged 
${\cal N}=8$ supergravity in five dimensions, hep-th/9812035.
\bibitem{KLM} A. Karch, D. Lust and A. Miemiec, Phys. Lett B454 (1999) 265,
hep-th/9901041.
\bibitem{FGPW1} D. Z. Freedman, S. S. Gubser, 
K. Pilch, and N. P. Warner,
Renormalization group flows from holography-supersymmetry and 
a c-theorem, hep-th/9904017.
\bibitem{GPPZ1} L. Girardello, M. Petrini, M. Porrati, and
A. Zaffaroni, JHEP 05 (1999) 026, hep-th/9903026.
\bibitem{AGPZ} D. Anselmi, L. Girardello, M. Porrati, and
A. Zaffaroni, A note on the 
holographic beta and $c$ functions,
hep-th/0002066.
\bibitem{GPPZ} L. Girardello, M. Petrini, M. Porrati, and
A. Zaffaroni, The supergravity dual 
of ${\cal N}=1$ Super Yang-Mills 
Theory, hep-th/9909047.
\bibitem{G0} S. S. Gubser, Non-conformal examples of AdS/CFT, 
hep-th/9910117.
\bibitem{BVV} J. de Boer, E. Verlinde and H. Verlinde, On the Holographic 
Renormalization Group, hep-th/9912012.
\bibitem{KN} I. Klebanov and N. Nekrasov, Gravity duals of fractional branes
and logarithmic RG flow, hep-th/9911096.  
\bibitem{KT}I. Klebanov and A. Tseytlin, Gravity Duals of 
Supersymmetric $SU(N)\times SU(N+M)$ Gauge Theories, 
hep-th/0002159.
\bibitem{KLT} P. Kraus, F.Larsen and S.P. Trivedi, JHEP 03 (1999) 003,
hep-th/9811120.
\bibitem{KW} I. R. Klebanov and E. Witten, Nucl. Phys.
B556 (1999) 89, hep-th/9905104.
\bibitem{FGPW2} D. Z. Freedman, S. S. Gubser, K. Pilch, and N. P. Warner,
Continuous distributions of D3-branes and 
gauged supergravity, 
hep-th/9906194.
\bibitem{BS} A. Brandhuber and K. Sfetsos, 
Wilson loops from multicentre and rotating branes,
mass gaps and phase structure 
in gauge theories, hep-th/9906201.
\bibitem{CGLP} M. Cvetic, S. S Gubser, H. Lu and C. Pope, Symmetric potentials
of gauged supergravities in diverse dimensions and Coulomb branch of 
gauge theories, hep-th/9909121.
\bibitem{G} S. S. Gubser, Curvature singularities: the good, the bad,
and the naked, hep-th/0002160 
\bibitem{DF} O. DeWolfe and D. Z. Freedman, 
Notes on fluctuations and correlation functions in holographic 
renormalization group flows, hep-th/0002226.
\bibitem{AF1} G. Arutyunov and S. Frolov, 
Nucl.Phys.B544 (1999) 576-589, hep-th/9806216.
\bibitem{aft2} G. Arutyunov, S. Theisen and S. Frolov, 
work in progress.
\bibitem{Osborn} H. Osborn, Nucl.Phys.B363 (1991) 486-526.
\bibitem{MR} P. Minces and V. O. Rivelles, 
Scalar field theory in the AdS/CFT correspondence 
revisited, hep-th/9907079.
\bibitem{AF2} G. Arutyunov and S. Frolov, Four-point Functions of
Lowest Weight CPOs in ${\cal N}=4$ SYM$_4$ in Supergravity
approximation, hep-th/0002170.

}
\end{thebibliography}
\end{document}